\documentstyle[aps,amsmath,eqsecnum]{revtex}

\begin{document}
\title{Renormalization group treatment of the scaling properties of
finite systems with subleading long-range interaction}
\author{H. Chamati$^1$\thanks{e-mail: chamati@issp.bas.bg} 
and D. M. Dantchev$^2$\thanks{e-mail: danield@bgcict.acad.bg}}
\address{
$^1$Institute of Solid State Physics - BAS,
Tzarigradsko chauss\'{e}e 72, 1784 Sofia, Bulgaria. \\
$^2$Institute of Mechanics - BAS, Acad. G. Bonchev St. bl. 4,
1113 Sofia, Bulgaria.}
\preprint{\today}
\maketitle
\begin{abstract}
The finite size behavior of the susceptibility, Binder cumulant
and some even moments of the magnetization of a fully finite
$O(n)$ cubic system of size $L$ are analyzed and the corresponding 
scaling functions
are derived within a field-theoretic $\varepsilon$-expansion
scheme under periodic boundary conditions. 
We suppose a van der Waals type long-range interaction
falling apart with the distance $r$ as $r^{-(d+\sigma)}$, where
$2<\sigma<4$, which does not change the short-range critical
exponents of the system. Despite that the system belongs to the
short-range universality class it is shown that above the bulk
critical temperature $T_c$ the finite-size corrections decay in a
power-in-$L$, and not in an exponential-in-$L$ law, which is
normally believed to be a characteristic feature for such systems.
\end{abstract}

\section{Introduction}
It is well known that the critical properties of a given bulk system
depend on a small number of parameters like its dimensionality, the
symmetry of the order parameter and the long-rangeness of the
interaction in the system under consideration. If the Fourier
transform of the interaction $v({\bbox q})$ has a small $|{\bbox q}|$
expansion of the form
\begin{equation}\label{expand2}
v({\bbox q}) = v_{0} + v_{2}{\bbox q}^{2} 
+v_{\sigma} {\bbox q}^{\sigma}+w({\bbox q}),
\end{equation}
with $w({\bbox q})/{\bbox q}^\sigma\rightarrow 0$ when ${\bbox q}\to0$ and
$\sigma\ge 2$, then the thermodynamic critical behavior of the
system is supposed to be like that of an entirely short-ranged
system \cite{fisher72}. In the opposite case, when $\sigma<2$ the
critical behavior differs essentially \cite{fisher72,suzuki73}
from that of the short-range system and is characterized by
critical exponents that do depend on $\sigma$ (below the
corresponding upper critical dimension that is $d_u=2\sigma$ in
this case)\cite{fisher72}. On the basis of the above bulk picture
one normally supposes that in the finite systems the same general
property will take place: if $\sigma\ge 2$ the finite-size
behavior will be that of the corresponding short-ranged
finite-size systems \cite{fisher86}, characterized by
exponentially fast decay of the finite-size dependence of the
thermodynamic quantities (at least when the critical region of the
system is leaved in the direction towards higher temperatures; the
low-temperature behavior depends on additional features like
existence, or not, of a spin-wave excitations - Goldstone bosons).
It turns out that the last is {\it not } true, at least for
$2<\sigma<4$, and an evidence about that within the framework of
the mean spherical model has been reported in \cite{dantchev01}.
For example, it has been demonstrated that the finite-size
dependence of the susceptibility in such a system is given by
($2<d<4$, $2<\sigma<4$, $d+\sigma<6$)
\begin{equation}
\chi(t,h;L)=L^{\gamma/\nu}Y(x_1,x_2,bL^{2-\sigma-\eta}),
\label{starting}
\end{equation}
or, equivalently,
\begin{equation}
\chi(t,h;L)=L^{\gamma/\nu}
\left[Y^{sr}(x_1,x_2) + bL^{2-\sigma-\eta} Y^{lr}(x_1,x_2)\right],
\end{equation}
where $x_1=c_1tL^{1/\nu}$, $x_2=c_2hL^{\Delta/\nu}$, and $Y$,
$Y^{sr}$  and $Y^{lr}$ are universal functions (recall that
$\eta=0$ for the short-range spherical model). The quantities
$c_1$, $c_2$ and $b$ are nonuniversal constants. In the
high-temperature, unordered phase, where
$tL^{1/\nu}\rightarrow\infty$, one observes \cite{dantchev01} that
the long-range portion of the interaction between spin degrees of
freedom gives rise to contributions of the order of
$bL^{-(d+\sigma)}$. In other words the {\it subleading} long-range
part of the interaction gives rise to a {\it dominant} finite-size
dependence in this regime which is governed by a {\it
power-in-$L$} law. More explicitly, one obtains $Y^{sr}(x_1,0)\sim
\exp(-{\rm const.} \ x_1^\nu)$, while
\begin{equation}\label{aY}
Y^{lr}(x_1,0)\sim x_1^{-d\nu-2\gamma},
\end{equation}
when $x_1\rightarrow\infty$ \cite{dantchev01}. This asymptotic is
supported from the existing both exact and perturbative  results for
models with  {\it leading} long-range interaction
\cite{singh89,brankov91,chamati01,chamati02}. Note that
(\ref{aY}) implies for the
temperature dependence of this corrections that
\begin{equation}
\chi(t,h;L)-\chi(t,h;\infty)\sim t^{-d\nu-2\gamma}L^{-(d+\sigma)},
\ tL^{1/\nu}\rightarrow\infty. \label{plc}
\end{equation}

In addition, let us note that the standard finite-size scaling
\cite{Fisher72,fisherbarber72,barber83,privman90,brankov00} is
usually formulated in terms of {\it only one} reference length,
namely the bulk correlation length $\xi$. The main statements of
the theory are that:

{\it i)} The only relevant variable in terms of which the properties
of the finite system depend in the neighbourhood of the bulk critical
temperature $T_c$ is $L/\xi$.

{\it ii)} The rounding of the phase transition in a given finite
system sets in when $L/\xi=O(1)$.

The tacit assumption is that all other reference lengths will lead
{\it only to corrections} towards the above picture. As it is
clear from Eqs. (\ref{starting}) - (\ref{plc}) this is not the
case in systems with subleading long-range interactions. This is
an important class of systems. It contains all nonpolar fluids
where the dominant interaction is supposed to be of van der Waals
type, i.e. of the type given by Eq. (\ref{expand2}) with
$d=\sigma=3$.

In fact a similar problem has been recently studied by Chen and
Dohm \cite{chendohm2,chendohm3,chendohm5}.
They considered a
field-theoretical model with short-range interactions and
wavelength-dependent cutoff of fluctuations $\Lambda$. They
observe corrections to the infinite system thermodynamic behavior
going as an inverse power law in $L$ that do depend also on
$L\Lambda$ and not only on $L/\xi$. As it has been clarified in
\cite{dantchev01} the power law contributions to the finite size
corrections result there from the interplay of two features of
that model. The first is a sharp cutoff of fluctuations in
momentum space and the second is the removal of all the terms
beyond the $q^2$ one  in (\ref{expand2}), which has the effect of
introducing an effective interaction that falls off as a power law
in the separation between degrees of freedom.  This power-law
interaction leads immediately to power-law contributions to the
finite size corrections.

Theoretically the critical properties of finite-size systems have
been studied on the examples of exactly solvable models, by
renormalization group calculations - both in the field-theoretical
framework and in the real space, by conformal invariance and by
numerical (mainly Monte Carlo) simulations. An essential part of
these investigations is well described in a series of reviews
\cite{barber83,privman90,brankov00,zinnjustin96,henkel99}.

The $O(n)$ models are the most often used examples on the basis of
which one studies the scaling properties of finite-size systems.
The  best investigated cases are those of the $n=1$ (Ising model)
and the limit $n=\infty$, which includes the spherical model
\cite{privman90,brankov00}. The last model is especially suitable
for the investigation of its finite-size properties since it is
exactly solvable for any $d$ even in the presence of an external
magnetic field $H$. For $n\ne1,\infty$ there are no exact results
and the preferable analytical method for the derivation of the
properties of the corresponding models (like $XY$, i.e. $n=2$, and
Heisenberg, i.e. $n=3$) is that one of the renormalization group
theory. An important amount of information for such systems is in
addition derived by numerical simulations, normally via Monte
Carlo methods. As a rule the investigations are concentrated on
interactions of finite range. As examples of long-range
interactions in addition to the equivalent neighbors the case of
power-law decaying interactions have been considered. In the case
of $\sigma<2$ analytically only the finite-size scaling properties
of the $n=\infty$ limit are well established. For finite $n$ a
limited number of recent numerical results
\cite{romano96,luijten97,bayong99,luijten99}, as well  as few
theoretical works \cite{chamati01,chamati02,luijten99,korucheva91} are
available. In \cite{korucheva91} one studies the crossover from
long to short-range interactions, i.e. the limit
$\sigma\rightarrow2$ and concludes that the renormalized values of
the temperature and the coupling constant are continuous functions
of $\sigma$. The case of $\sigma<2$ has been investigated in
references \cite{chamati01,chamati02,luijten99} (under periodic boundary
conditions). It has been found that, as for the bulk systems
\cite{fisher72,yamazaki77}, the critical behavior depends on the
small parameter $\varepsilon=2\sigma-d$, where $2\sigma$
corresponds to the upper critical dimension in such systems
\cite{fisher72,yamazaki77}. The results are obtained in powers of
$\sqrt{\varepsilon}$. The quantities of interest have been the
shift of the critical coupling, the susceptibility and the Binder
cumulant $B$ at the critical temperature $T_c$
\cite{luijten99,chamati01}
and above it \cite{chamati01,chamati02} as a function of $\varepsilon$. It
has been found that the numerical results obtained in
\cite{luijten99} for the Ising model do not agree with the
predicted (up to one loop order) behavior of $B$
\cite{chamati01,luijten99}. In \cite{bayong99} one even reports
disagreement with the well established theoretically fact that the
critical exponent of the system do not depend on $\sigma$ if
$\sigma>2$. In \cite{romano96} a Heisenberg model and in
\cite{luijten97} an Ising model have been investigated by Monte
Carlo methods in the case when their critical behavior is
characterized by classical (mean-field) critical exponents.

In the present article we will consider the case of long-range
power-law decaying interaction characterised by $\sigma>2$ in its
Fourier transform. As it was already mentioned above the recently
obtained results for $n=\infty$ limit indicates that the well-spread
opinion that such an interaction is uninteresting for the
critical behavior of the finite system \cite{fisher86} is not fully
correct. Here, following the method used in \cite{chamati01} we will
generalize the results available for $n=\infty$ to the case of finite
$n$. We will use $\varepsilon$-expansion technique up to one loop
order in the interaction coupling. We will investigate the behavior of
the Binder cumulant, susceptibility, and some more general even
moments of the order parameter.

The plan of the article is as follows. In Section~\ref{gencon} we
review, briefly, the $\varphi^4$-model with long-range interaction
and discuss its bulk critical behavior. Section~\ref{secfss} is
devoted to the explanation of the methods used here to achieve our
analysis. We end the section with the computation of some
thermodynamic quantities of interest. In Section~\ref{discussion}
we discuss our results briefly. In the remainder of the paper we
present details of the calculations of some formula used
throughout the paper.

\section{General considerations}\label{gencon}
In the vicinity of its critical point the Heisenberg model, with
short as well as long-range interaction decaying in a power-law,
is equivalent to the $d$-dimensional ${\cal O}(n)$-symmetric model
\begin{equation}\label{model}
\beta{\cal H}\left\{\varphi\right\}=\frac12\int_V d^d{\bbox x}\left[
\left(\nabla\varphi\right)^2+b\left(\nabla^{\sigma/2}\varphi\right)^2
+r_0\varphi^2+\frac12u_0\varphi^4\right],
\end{equation}
where $\varphi$ is a short-hand notation for the space dependent
$n$-component field $\varphi(x)$, $r_0=r_{0c}+t_0$ ($t_0\propto
T-T_c$) and $u_0$ are model constants. $V$ is the volume of the
system. In equation~(\ref{model}), we assumed $\hbar=k_B=1$ and
the size scale is measured in units in which the velocity of
excitations $c=1$. We note that the second term in the model
denotes ${\bbox q}^\sigma|\varphi({\bbox q})|^2$ in the momentum
representation where the parameter $\sigma>0$ (with $\sigma/2$
being noninteger) takes into account the contribution of the
long-range interactions in the system. In (\ref{model}) $\beta$ is
the inverse temperature. Here we will consider periodic boundary
conditions. This means
\begin{equation}
\varphi({\bbox x})=L^{-d}\sum_{\bbox q}\varphi({\bbox q})
\exp\left(i{\bbox q}\cdot {\bbox x}\right),
\end{equation}
where ${\bbox q}$ is a discrete vector with components $q_i=2\pi
n_i/L$ $\left(n_i=0,\pm1,\pm2,\cdots,\ i=1,\cdots,d\right)$ and a
cutoff $\Lambda\sim a^{-1}$ ($a$ is the lattice spacing). In this
paper we are interested in the continuum limit, i.e. $a \to 0$. As
long as the system is finite we have to take into account the
following assumptions $L/a\to\infty$, $\xi\to\infty$ while $\xi/L$
is finite.

The Hamiltonian (\ref{model}) is, of course, well known in the
literature. First, it has been used to investigate the critical
behaviour of systems with reduced space dimensionality exhibiting
phase transitions \cite{fisher72}. Let us recall that in such
systems a phase transition can occur only if the interaction is
long-ranged enough. The critical behaviour of the model depends
strongly upon the nature of the interaction controlled by the
parameter $\sigma$. With $\sigma\le 2$ it has been used for
detailed investigation of the critical behavior of $O(n)$ models
including questions like the $\sigma$, $d$ and $n$ dependence of
the critical exponents and critical amplitude ratios, as well as
for calculation of their values, and for determining of the
universal scaling functions of both the infinite, as well as of
finite systems. In this case the critical exponents of the system
are $\sigma$ dependent. By increasing $\sigma$, a crossover from
long-range critical behavior to short-range one takes place. The
crossover happens at a point, which can be determined from general
considerations (see for example page 71 of reference
\cite{cardy96}). This `critical' value of $\sigma$ is given by
$\sigma=2-\eta$, where $\eta$ is the Fisher exponent for the short
range model. When $\sigma>2$ one usually considers the model as
equivalent to $\sigma=2$ case and omits the
$b\left(\nabla^{\sigma/2}\varphi\right)^2$ term in the
Hamiltonian, since it was believed that this term does not
contribute to the critical behavior of the system. Indeed, in this
case, the critical exponents does not depend on the parameter
$\sigma$. As it was already mentioned, such a procedure can lead
to incorrect results for finite-size systems. This was
demonstrated in \cite{dantchev01} on the example of $n=\infty$
model. In the current article we will demonstrate that the same
remains true also for a finite $n$.

The investigation of the critical phenomena of the model
(\ref{model}) for the case $\sigma>2$ is achieved by considering
the long range interaction as a perturbation to the short range
one \cite{sak73,yamazaki80,honkonen89}. This allows the adaptation
of the theory of Feynman diagrams to systems with subleading long
range interaction. As a consequence the upper critical dimension
remains unchanged by that interaction and the critical exponents
are those of the model with pure short range interaction. The
interested reader can find more details in refrences
\cite{sak73,yamazaki80,honkonen89}.

Before starting to explore the scaling properties of the field
theoretical model (\ref{model}) confined to a finite geometry and
under periodic boundary conditions, we will give a brief heuristic
derivation of the finite scaling hypothesis, based on the idea of
renormalization group. Here we are interested in the continuum
limit when the lattice spacing completely disappears. Using
dimensional regularization the integrations over wave vectors of
the fluctuations are convergent and are evaluated without cutoff.
When some dimensions of the system are finite the integrals over
the corresponding momenta are transformed into sums. Since the
lattice spacing is taken to be zero, the limits of the sums still
extend to infinity.

>From general renormalization group considerations a
multiplicatively renormalizable observable $X$, the susceptibility
for example, will scale like
\begin{equation}\label{chi}
X[t,g,b,\mu,L]=\zeta(\rho)X\left[t(\rho),g(\rho),b(\rho),\mu\rho,L\right],
\end{equation}
where $t=(T-T_c)/T_c$ is the reduced temperature, $g$ is a
dimensionless coupling constant and $L$ is the finite-size scale.
The length scale $\mu$ is introduced in order to control the
renormalization procedure. Here $b(\rho)$ is an irrelevant from RG
point of view variable which mimics the influence of the
subleading long range interaction on the critical behavior of the
system. Equation (\ref{chi}) is obtained using the assumption that
the size $L$ of the system does not renormalize \cite{brezin82}.

It is known (see, e.g., \cite{brezin82}) that in the bulk limit,
when $g(\rho)$ approaches its stable short-range fixed point $g^*$
of the theory, then we have
\begin{equation}\label{rho}
t(\rho)\approx t\rho^{1/\nu-2}, \ \ \ \
\zeta(\rho)\approx\rho^{\gamma_x/\nu-p_x}
\ \ \ \ {\rm and} \ \ \ \ b(\rho)\approx b\rho^{\gamma_b-2},
\end{equation}
where $\gamma_x$ and $\nu$ are the bulk critical exponents
measuring the divergence of the observable $X$ and the correlation
length, respectively, in the vicinity of the critical point and
$\rho$ is a scaling parameter. The exponent $p_x$ is the dimension
of the observable $X$, defined in equation (\ref{chi}). The critical
exponent $\gamma_b=2-\eta-\sigma$ \cite{yamazaki77a}. Using
dimensional analysis together with equation~(\ref{chi}) one gets
\begin{equation}\label{fss}
X[t,g,b,\mu,L]=(\mu\rho)^{p_x}\zeta(\rho) X\left[t(\rho)(\rho
\mu)^2, g(\rho),b(\rho)(\rho\mu)^{2},1,L/\mu\rho\right].
\end{equation}
Choosing the arbitrary parameter $\rho=L/\mu$, we obtain
our final result for the scaling form of an
observable $X$ in the case, when there are subleading long-range
interaction in the finite system
\begin{equation}\label{universal}
X[t,g,b,\mu,L]=L^{\gamma_x/\nu}f\left(tL^{1/\nu},bL^{2-\sigma-\eta}\right).
\end{equation}
Here the function $f(x)$ is a universal function of its argument.
Note that equation (\ref{universal}) is the analog, for
finite system, of the result obtained in \cite{yamazaki77a}. In
the remainder of this paper we will verify the scaling relation
(\ref{universal}) in the framework of model (\ref{model}).

\section{finite-size analysis}\label{secfss}
The method we will adopt here is widely used in the exploration of
the scaling properties of finite systems in the vicinity of their
critical point. It is based on the idea of using a mode expansion,
i.e. one treats the zero mode of the order parameter, which is
equivalent to the magnetization, separately from the higher modes
($q\neq0$). The nonzero modes are treated perturbatively in
combination with the loop expansion. The finite modes are traced
over to yield an effective Hamiltonian for the zero mode:
\begin{equation}
\exp\left[-{\cal H}_{\rm eff}\right]={\rm Tr}_{\phi_{q\neq0}}
\exp\left[-{\cal H} \left(\phi_{q=0}, \phi_{q\neq0}\right)\right].
\end{equation}
After performing this operation one ends up with an effective
Hamiltonian of the form~(see Appendix \ref{effham})
\begin{equation}\label{effective}
{\cal H}_{\rm eff}=\frac12 L^d\left(R\phi^2+\frac12U\phi^4\right),
\end{equation}
where the effective coupling constants are given by
\begin{mathletters}
\begin{eqnarray}
R&=&r_0+(n+2)u_0 L^{-d}\sum_{{\bbox q}\neq0}\frac1{r_0+{\bbox q}^2
+b|{\bbox q}|^\sigma},\\
U&=&u_0-(n+8)u_0^2L^{-d}\sum_{{\bbox q}\neq0}\frac1{\left(r_0+{\bbox q}^2
+b|{\bbox q}|^\sigma\right)^2}.
\end{eqnarray}
\end{mathletters}
In the remainder of this paper we will compute, to the lowest order in
$\varepsilon=4-d$, the effective coupling constants, with the initial
coupling constants renormalized as in their bulk critical theory,
since it has been shown that to the one loop order the
renormalization of the finite theory is a consequence of the
renormalization of the bulk one \cite{brezin82}.

Simple dimensional analysis shows that the effective coupling
constants should have the following scaling forms:
\begin{equation}\label{scalingcoupling}
R=L^{\eta-2}f_R\left(tL^{1/\nu},bL^{2-\sigma-\eta}\right)
\ \ \ \ \text{and} \ \ \ \
U=L^{d-4+2\eta}f_U\left(tL^{1/\nu},bL^{2-\sigma-\eta}\right),
\end{equation}
for $t\gtrsim0$, where $f_R$ and $f_U$ are scaling functions which
are properties of the bulk critical point. They are analytic at $t=0$.
This is a consequence of the fact that only finite modes have been
integrated out.

After evaluating the explicit forms of the functions $f_R$ and $f_U$,
we can deduce results for the different thermodynamic quantities and
the expressions of their respective scaling functions.

In order to investigate the large scale physics of the finite
system, one has to calculate thermal averages with respect to the
new effective Hamiltonian defined in~(\ref{effective}). They are
related to the thermodynamic functions of the system under
consideration. The averages of the field $\phi$ are defined by
\begin{equation}\label{averages}
{\cal M}_{2p}=\left<\left(\phi^2\right)^p \right> =\frac{\int
d^n\phi \ \phi^{2p}\exp\left(-{\cal H}_{\rm eff}\right)} {\int
d^n\phi \exp\left(-{\cal H}_{\rm eff}\right)}.
\end{equation}
With the aid of the appropriate rescaling
$\Phi=\left(UL^d\right)^{1/4}\phi$, we can transform the effective
Hamiltonian into
\begin{equation}\label{neweff}
{\cal H}_{\rm eff}=\frac12z \Phi^2+\frac14\Phi^4,
\end{equation}
where the `scaling variable' $z=RL^{d/2}U^{-1/2}$ is an important
quantity which has been used in many occasions in the
investigations of finite-size scaling in critical systems (see for
example references \cite{brezin85,rudnick85}). Explicit
expressions for some thermodynamic averages of the type
(\ref{averages}) as well as their asymptotics are presented in
Appendix \ref{momenta}.

With the effective Hamiltonian (\ref{neweff}), we obtain the general
scaling relation
\begin{equation}\label{scaling}
{\cal M}_{2p}=L^{-p(d-2+\eta)}\frac{L^{p(d-4+2\eta)/2}}{U^{p/2}}
f_{2p}\left(RL^{2-\eta} \frac{L^{(d-4+2\eta)/2}}{U^{1/2}}\right)
\end{equation}
for the momenta of the field $\phi$. Having in mind Eqs.
(\ref{scalingcoupling}), we can write down Eq. (\ref{scaling}) in the
following scaling form
\begin{equation}\label{Scaling}
{\cal M}_{2p}=L^{-p(d-2+\eta)}{\cal F}_{2p}(tL^{1/\nu},bL^{2-\sigma-\eta}),
\end{equation}
in agreement with the finite-size scaling predictions of
(\ref{universal}). In equation (\ref{Scaling}), the functions
${\cal F}_{2p}(x)$ are universal.

All the measurable thermodynamic quantities can be obtained from the
momenta ${\cal M}_{2p}$. For example the susceptibility is obtained
from
\begin{equation}\label{chi1}
\chi=\frac1n\int_V d^dx\left<\varphi({\bf
x})\varphi(\bf{0})\right> =L^{2-\eta}{\cal
F}_2(tL^{1/\nu},bL^{2-\sigma-\eta}).
\end{equation}
Another quantity of importance for numerical analysis of the finite-size
scaling theory is the Binder's cumulant defined by
\begin{equation}\label{binder}
{\cal B}=1-\frac13\frac{{\cal M}_4}{{\cal M}_2^2}.
\end{equation}

In the remainder of this section we concentrate on the computation of
the coupling constants $R$ and $U$ of the effective Hamiltonian
(\ref{effective}) for the system with subleading long-range
interaction decaying
with the distance as a power law. As a consequence we will deduce
results for the characteristic variable
$z=RU^{-1/2}L^{2-\eta-\varepsilon/2}$, the susceptibility $\chi$ and the
Binder's cumulant ${\cal B}$.

\subsection{Computation of the effective coupling constants}
The finite-size corrections to the coupling constant $r_0$ in the
mode expansion reads
\begin{equation}\label{Roneloop}
R=r_0+(n+2)u_0 L^{-d}\sum_{{\bbox q}\neq0}\frac1{r_0+{\bbox q}^2
+b|{\bbox q}|^\sigma}.
\end{equation}

One of the delicate problems in the finite size-scaling theory is
the analysis of the sums appearing in the mathematical equations,
which forms the basis of the investigation of the scaling as well
as thermodynamic properties of the system under consideration. In
our case this means that we have to find a way to evaluate the sum
appearing in the right hand side of (\ref{Roneloop}). In the
absence of the long-range interaction term ($b=0$) several methods
have been developed in order to investigate this sum. When
$b\neq0$, i.e. in the presence of the non analytic term in ${\bbox
q}$, a step towards the solution of this problem has been made in
reference~\cite{dantchev01}. It is based upon the idea that  in
the long distance physics one retains only those contribution to
the behavior of the quantities involved that are associated with
the effects of long-range fluctuations. In other words we will
consider the leading behavior that is due to the small ${\bbox q}$
contributions. Expanding in ${\bbox q}$, we obtain
\begin{equation}
R=r_0+(n+2)u_0 S_L(d,r_0,2)-(n+2)u_0b\left(1+
r_0\frac\partial{\partial r_0}\right)S_L(d,r_0,\sigma),
\end{equation}
where
\begin{equation}\label{sdef}
S_L(d,r,\sigma)=\frac1{L^d}\sum_{{\bbox q}\neq0}
\frac{|{\bbox q}|^{\sigma-2}}{r+{\bbox q}^2}.
\end{equation}
In order to evaluate the finite-size corrections to the bulk system
we have to analyze the finite-size behavior of the function
$S_L(d,r,\sigma)$. This is achieved by making use of the identity
\begin{equation}
\frac{q^{2p}}{r+q^2}=\int_{0}^{\infty}\exp[-(q^2 +r)t]
t^{-p}\gamma^*(-p,-rt) dt, \ \  p<1, \label{si}
\end{equation}
where $\gamma^*(a,x)$ is a single-valued analytic function of $a$ and
$x$, possessing no finite singularities \cite{abramovitz70}
\begin{equation}
\gamma^*(a,x)=e^{-x}\sum_{n=0}^{\infty}\frac{x^n}{\Gamma(a+n+1)}=
\frac{1}{\Gamma(a)}\sum_{n=0}^{\infty}\frac{(-x)^n}{(a+n)n!}, \ \
|x|<\infty.
\label{gammastar}
\end{equation}

Identity (\ref{si}) can be proven by integrating by parts the
corresponding series representations of $\gamma^*$. Similar
identity has been used in \cite{korucheva91} for the investigation
of the finite-size behavior of ${\cal O}(n)$ model system with a
crossover from leading long-range interaction to the short-range
case, i.e. $\sigma\to2^-$.

With the help of this identity one obtains ($0\le p<1$)
\begin{equation}\label{sresult}
S_L(d,r,2(p+1))
=\frac1{(2\pi)^d}\int d{\bbox q}\frac{{\bbox q}^{2p}}
{r+{\bbox q}^2}+L^{2-d-2p}I_{{\rm
scaling}}^p(rL^2,d),
\end{equation}
where
\begin{equation}\label{Iscaling}
I_{\rm scaling}^p(x,d)=(4\pi)^{p-1}\int_0^\infty e^{-x\frac u{4\pi^2}}
u^{-p}\gamma^*(-p,-\frac x{4\pi^2}u)
\left[{\cal A}^d(u) -\left(\frac\pi u\right)^{d/2}-1 \right]du,
\end{equation}
with
$$
{\cal A}(u)=\sum_{k=-\infty}^{\infty}e^{-k^2u}=\sqrt{\frac\pi u}
{\cal A}\left(\frac{\pi^2}u\right).
$$

Finally for the effective coupling constant $R$ we obtain
\begin{eqnarray}
R&=&r_0+\frac{(n+2)u_0}{(2\pi)^d}\left(\int\frac{d{\bbox
q}}{r_0+{\bbox q}^2} -b\int d{\bbox q}\frac{|{\bbox q}|^\sigma}
{(r_0+{\bbox q}^2)^2}\right)
+(n+2)u_0L^{2-d}I^0_{\rm scaling}\left(r_0L^2,d\right)\nonumber\\
& &-(n+2)u_0b\left(1+r_0\frac\partial{\partial
r_0}\right)L^{4-d-\sigma}I^{\frac{\sigma-2}2}_{\rm
scaling}\left(r_0L^2,d\right),
\end{eqnarray}

Now, we renormalize the theory by introducing the field
theoretical renormalization constants, i.e. the scale field
amplitude $Z$, the coupling constant renormalization $Z_g$, and
$Z_t$ -- renormalizing the $\varphi^2$ insertions in the critical
theory. This allows to replace the model bare constants $r_0$ and
$u_0$ in the last equation by their renormalized counterparts
trough:
\begin{equation}\label{definition}
t=Z Z_t^{-1}(r_0-r_{0c}) \ \ \ \ \text{and } \ \ \ \
g=\mu^{-\varepsilon}Z^2Z_g^{-1}u_0S_d^{-1},
\end{equation}
where $\mu$ is a renormalization scale, which will be set equal to
1, $S_d=\frac12(4\pi)^{d/2}\Gamma(d/2)$ is a phase space factor
and \cite{yamazaki80,honkonen89}
\begin{mathletters}\label{renormalization}
\begin{eqnarray}
Z&=&1+{\cal O}(g^2),\\
Z_t&=&1+\frac{n+2}\varepsilon g+{\cal O}(g^2),\\
Z_g&=&1+\frac{n+8}\varepsilon g+{\cal O}(g^2),
\end{eqnarray}
are the usual renormalization amplitudes to one-loop order.
\end{mathletters}

Finally, using dimensional regularization, at the fixed point
$g^*=\frac\varepsilon{n+8}+{\cal O}(\varepsilon^2)$ of the
theory, in the case $d+\sigma<6$ we obtain
\begin{eqnarray}\label{finalR}
RL^2&=&y\left(1+\frac\varepsilon2\frac{n+2}{n+8}\ln y\right)+
\frac\varepsilon4\frac{n+2}{n+8}bL^{2-\sigma}\frac{(2+\sigma)\pi}
{\sin\left(\pi\frac\sigma2\right)}y^{\sigma/2}+\varepsilon
\frac{n+2}{n+8} S_4
I^0_{\rm scaling}\left(y,4\right)\nonumber\\
&&-\varepsilon S_4\frac{n+2}{n+8}bL^{2-\sigma}
\left(1+y\frac\partial{\partial
y}\right)I^{\frac{\sigma-2}2}_{\rm scaling}\left(y,4\right),
\end{eqnarray}
where we have introduced the scaling variable $y=tL^{1/\nu}$ with
$\nu^{-1}=2-\frac{n+2}{n+8}\varepsilon+{\cal O}
\left(\varepsilon^2\right)$. Equation (\ref{finalR}) shows that
the effective coupling constant $R$ has the scaling form predicted
in equation (\ref{scalingcoupling}). At this order, the exponent
$\eta=0$, and verifying the powers of $\eta$ in the this
expression requires a higher order computation. In the particular
case $b=0$ from equation (\ref{finalR}) we recover the result of
reference \cite{brezin85}.

When the system under consideration is confined to a finite
geometry, instead of the coupling constant $u_0$, we have the
shifted effective coupling constant $U$ given by:
\begin{equation}\label{Uoneloop}
U=u_0-(n+8)u_0^2L^{-d}\sum_{{\bbox q}\neq0}\frac1{\left(r_0+{\bbox q}^2
+b|{\bbox q}|^\sigma\right)^2}.
\end{equation}
Remark that the summand in the right hand side can be expressed as
the derivative of the summand in right hand side of
Eq.~(\ref{Roneloop}) with respect to $r_0$. Consequently the result
for the effective coupling constant $U$ can be derived easily from
that of $R$. Using that observation one gets
\begin{eqnarray}\label{finalU}
UL^{\varepsilon}&=&\frac\varepsilon{n+8}S_4
\left(1+\frac\varepsilon2(1+\ln y)\right)+
\frac{\varepsilon^2}{n+8}\frac{bL^{2-\sigma}}{8}
S_4\frac{\sigma\pi(\sigma+2)}
{\sin\left(\pi\frac \sigma2\right)}y^{\frac\sigma2-1}\nonumber\\
&&+\frac{\varepsilon^2}{n+8}S_4^2\frac\partial{\partial
y}I^0_{\rm scaling}\left(y,4\right)
-\frac{\varepsilon^2}{n+8}S_4^2 bL^{2-\sigma}
\left(2\frac\partial{\partial y}+
\frac{\partial^2}{\partial y^2}\right)I^{\frac{\sigma-2}2}_{\rm
scaling}\left(y,4\right),
\end{eqnarray}
at the fixed point, in agreement with the scaling relations of
equation (\ref{scalingcoupling}). Equation (\ref{finalU}) generalizes
the resuls of refrence \cite{brezin85} to the case when subleading
long range interaction is taken into account.

Note that the effective coupling constant $U$ has a finite limit
at the critical point, i.e. in the limit $t\to0$. Indeed, as the
reduced temperature vanishes it is possible to use the expansion
\begin{equation}\label{smally}
I^0_{\rm scaling}\left(y,4\right)=I^0_{\rm scaling}\left(0,4\right)+
\frac{S_4^{-1}}2y\left({\cal C}-\ln y\right)+{\cal O}(y^2),
\end{equation}
where
\begin{equation}
{\cal C}=\int_0^\infty\frac{du}u
\left[\exp\left(-\frac{u}{4\pi^2}\right)
-\frac{u^2}{\pi^2}{\cal A}^{4}(u)+
\frac{u^2}{\pi^2}\right]=2.2064... .
\end{equation}
After substitution of~(\ref{smally}) in~(\ref{finalU}) the terms
proportional to $\log y$ cancel, which shows that the coupling
constant $U$ is finite at $t=0$. Whence, one gets (for $y\to0$)
\begin{equation}
UL^{\varepsilon}=\frac\varepsilon{n+8}S_4
\left(1+\frac\varepsilon2{\cal C}\right)
-\frac{\varepsilon^2}{n+8}S_4^2 bL^{2-\sigma}
\left(2\frac\partial{\partial y}+
\left.\frac{\partial^2}{\partial y^2}\right)I^{\frac{\sigma-2}2}_{\rm
scaling}\left(y,4\right)\right|_{y=0},
\end{equation}
showing that it is possible to evaluate $U$ at the critical point, i.e.
it is safe now to set $y=0$.


\subsection{Some thermodynamic quantities}
\subsubsection{Binder's cumulant}
In this subsection we are interested in the evaluation of the
Binder's cumulant ratio, which plays a fundamental role in the
investigation of the finite-size scaling theory by numerical
means. Here we will give only the analytical expressions.
Unfortunately there are no numerical simulation which can approve
or not the results we obtain throughout this paper.

Close to the critical point, i.e. in the region $tL^{1/\nu}\ll1$,
we obtain for for the Binder's cumulant ratio
\begin{eqnarray}\label{amplitude}
{\cal B}&=&1-\frac n{12}\frac{\Gamma^2\left[\frac14n\right]}
{\Gamma^2\left[\frac14(n+2)\right]}\left\{1-
z\left(\frac{\Gamma\left[\frac14(n+6)\right]}
{\Gamma\left[\frac13(n+4)\right]}+
\frac{\Gamma\left[\frac14(n+2)\right]}
{\Gamma\left[\frac14n\right]}
-2\frac{\Gamma\left[\frac14(n+4)\right]}
{\Gamma\left[\frac14(n+2)\right]}\right)\right.\nonumber\\
&&\left.+z^2\left(\frac{\Gamma\left[\frac14(n+6)\right]
\Gamma\left[\frac14(n+2)\right]}
{\Gamma\left[\frac14(n+4)\right]\Gamma\left[\frac14n\right]}
+3\frac{\Gamma^2\left[\frac14(n+4)\right]}
{\Gamma^2\left[\frac14(n+2)\right]}-n-1\right) +{\cal
O}\left(z^3\right)\right\},
\end{eqnarray}

The cumulant ${\cal B}$ is a function of the variable $z$, which is
itself a function of the scaling variable $y$. So, a knowledge of a
final expression for the function $z$, which appears in the all
thermodynamic quatities, is enough to deduce the value of the
Binder's Cumulant. At the fixed point, we obtain (for $y<<1$)
\begin{eqnarray}\label{zstar}
z^*(y)&\equiv& \left.\frac{R L^{2}}{\sqrt{UL^\varepsilon}}
\right|_{\rm fixed point}\nonumber\\
&=&\sqrt{\frac{n+8}{\varepsilon S_4}}\left[y-y\frac\varepsilon4\left(1-
\frac{n-4}{n+8}\ln y\right)+\frac{3n}{n+8}\frac\varepsilon{16}
\frac{(2+\sigma)\pi}
{\sin\left(\pi\frac d\sigma\right)}y^{\sigma/2}\right.\nonumber\\
&&+\frac{n+2}{n+8}\varepsilon S_4\left(I^0_{\rm scaling}
\left(y,4\right)-bL^{2-\sigma}\left(1+y\frac\partial{\partial y}\right)
I^{\frac{\sigma-2}2}_{\rm scaling}\left(y,4\right)\right)\nonumber\\
&&\left.-\frac12\varepsilon S_4y\left(\frac\partial{\partial
y}I^0_{\rm scaling}\left(y,4\right)-bL^{2-\sigma}
\left(2\frac\partial{\partial y}+
\frac{\partial^2}{\partial y^2}\right)I^{\frac{\sigma-2}2}_{\rm
scaling}\left(y,4\right)\right)\right].
\end{eqnarray}
This expression shows that the Binder's Cumulant ${\cal B}$ has the 
required scaling form.
 At the critical point, i.e. at $y=0$,
we get
\begin{equation}\label{zfinal}
z^*(0)=-\sqrt{\varepsilon}\frac{4\sqrt2}\pi\frac{n+2}{\sqrt{n+8}}
\left[\ln2+bL^{2-\sigma}(2\pi)^{\sigma-2}(1-4^{\frac\sigma2-1})
\zeta\left(1-\frac\sigma2\right)
\zeta\left(2-\frac\sigma2\right)\right].
\end{equation}
This result is obtained with the help of the formula \cite{chamati001}
\begin{equation}
\int_0^\infty duu^{1-\nu}\left[{\cal A}^4(u)-1-\left(\frac\pi u
\right)^2\right]=8(1-4^{1-\nu})\pi^{2(1-\nu)}\Gamma(\nu)\zeta(\nu-1)
\zeta(\nu), \ \ \nu\neq0,2.
\end{equation}

Equation (\ref{zfinal}) is a generalization of the result of 
\cite{brezin85} obtained for the model with pure short range forces. 
Note that the form of the expansion in terms $\sqrt\varepsilon$ is 
kept but the coefficient is altered and now it is a function the 
parameter $\sigma$ controlling the long-range interaction.

Now we turn our attention to the behavior of Binder's cumulant ratio in the
limit $z\gg1$. In this case we obtain
\begin{equation}
{\cal B}=1-\frac13\left(1+\frac2n\right)\left[1-\frac2{z^2}+ {\cal
O}\left(\frac1{z^4}\right)\right],
\end{equation}
wherefrom one has ${\cal B}_n(\infty)=\frac23(1-1/n)$. This result
corresponds to a $n$-dimensional Gaussian distribution for $n$
independent components $\Phi_1, \cdots, \Phi_n$ of the vector
variable $\Phi$. For such a distribution it is easy to show that
$M_2=n<\Phi_i^2>$, and $M_4=n<\Phi_i^4>+n(n-1)<\Phi_i^2>^2$,
where $\Phi_i$ is any of the components of the vector $\Phi$, and
$<\cdots>$ means average with respect to one-component Gaussian
distribution ${\cal G}_1$. Having in mind that for ${\cal G}_1$
$<\Phi_i^4>=3<\Phi_i^2>^2$, one directly obtains that
$B_n=\frac23(1-1/n)$, in a full agreement with the above  renormalization
group result. Obviously, all limiting values lie in the interval from
${\cal B}=0$ (Ising model, $n=1$) to ${\cal B}=2/3$ (spherical model,
$n=\infty$).

\subsubsection{Magnetic susceptibility}
The system we consider here is confined to a fully finite
geometry. In this case it cannot exhibit a true phase
transition, i.e. the thermodynamic functions are not singular. In
the vicinity of the critical temperature, which corresponds to the
region $y\ll1$, the susceptibility behaves like
\begin{eqnarray}\label{sucyto0}
\chi&=&\frac2n\frac{L^2}{\sqrt{UL^\varepsilon}}
\frac{\Gamma\left[\frac14(n+2)\right]}
{\Gamma\left[\frac n4\right]}\left\{1-z\left(
\frac n4\frac{\Gamma\left[\frac n4\right]}
{\Gamma\left[\frac14(n+2)\right]}
-\frac{\Gamma\left[\frac14(n+2)\right]}
{\Gamma\left[\frac n4\right]}\right)\right.\nonumber\\
&&\left.+z^2\left(\frac{1-n}4+\frac{\Gamma^2\left[
\frac14(n+2)\right]}{\Gamma^2\left[\frac n4\right]}\right)
+{\cal O}(z^3)\right\}.
\end{eqnarray}
The susceptibility in this case is analytic at $t=0$. This is a
consequence of the analyticity of the effective coupling constants
$R$ and $U$. In order to get the final result for the susceptibility
one has to replace $R$ and $U$ by their respective expressions. After
performing this we find that $\chi$ has an expansion in powers of
$\sqrt\varepsilon$. This results is valid as long as we are concerned
by the case $d+\sigma<6$. Once we have $d+\sigma=6$, a $\ln L$ will
appear in the expression of the susceptibility. The source of this $\ln
L$ is coming from equation (\ref{udsigma6}) for the coupling constant
$U$ at the critical point (see Appendix \ref{ds6}). This is an extension
to finite $n$, by means
of perturbation method, of the
result obtained in reference \cite{dantchev01} for the spherical model.

In the region corresponding to $y\gg1$, we have
\begin{equation}
\chi=\frac1R\left[1-\frac{n+2}{z^2}+
{\cal O}\left(\frac1{z^3}\right)\right].
\end{equation}
Substituting the effective coupling constants $R$ and $U$ by their
respective expressions from equations (\ref{finalR}) and
(\ref{finalU}), and using the asymptotic expansions of 
$I^{p}_{\rm scaling}$ derived in
\cite{dantchev01},
we get
\begin{equation}\label{largey}
\chi=\chi_\infty\left[1-\varepsilon\frac{n+2}{n+8}\frac{S_4}yb
L^{2-\sigma}\left(C_{4,\frac{\sigma-2}2}y^{-2}-\frac{y^{\sigma/2}}
{4S_4}\frac{\sigma+2}{\sin{\pi\sigma/2}}\right)\right],
\end{equation}
where
\begin{equation}\label{Cpdef}
C_{d,p}=-\frac{(1+p)
4^{1+p}}{\pi^{d/2}}\frac{\Gamma(1+p+d/2)}{\Gamma(-p)}\sum_{{\bbox k}
\neq 0}\frac{1}{k^{d+2(p+1)}}.
\end{equation}
Expression (\ref{largey}) for the susceptibility shows that it has
the form given by the scaling hypothesis (\ref{chi1}). It demonstrates
also that in this regime the critical properties of the system are dominated by
the bulk critical behavior, with finite-size corrections in powers
of $L$.

\section{discussion}\label{discussion}
In the present article we have investigated the finite-size
scaling behavior of a fully finite $O(n)$ system with periodic
boundary conditions and in the presence of a long-range
interaction that does not alter the short-range  exponents of
critical its critical behavior.  The small $|{\bbox q}|$ expansion of the
Fourier transform of the interaction $v(\bbox q)$ is supposed to
be of the form
\begin{equation}
v(\vec{q}) = v_{0} + v_{2}{\bbox q}^{2} 
+v_{\sigma} {\bbox q}^{\sigma}+w({\bbox q}),
\end{equation}
with $w({\bbox q})\rightarrow0/{\bbox q}^\sigma$, 
when ${\bbox q}\to 0$ and $2<
\sigma < 4$. In the real $d$-dimensional space one can think about
interactions decaying as $r^{-(d+\sigma)}$. This is an important
class of interactions that include also van der Waals type
interactions.

For such a system, in the present article we have demonstrated that all the
even moments of the magnetization ${\cal M}_{2p}$, including the
susceptibility, can be written in the form
\begin{equation}
{\cal M}_{2p}=L^{-p(d-2+\eta)}{\cal
F}_{2p}(tL^{1/\nu},bL^{2-\sigma-\eta}),
\end{equation}
(see Eqs. (\ref{evenm}), (\ref{finalR}), (\ref{finalU}),
(\ref{sucyto0}), (\ref{largey})). Note that one has two scaling
variables needed in order to describe in a proper way the finite
size behavior of these quantities. A special  attention has been paid to two 
important quantities: the Binder's Cumulant and the susceptibility.

In the region $tL^{1/\nu}\gg1$ away from the critical point we
obtained for the Binder's cumulant ratio the expression
\begin{equation}
{\cal B}=1-\frac13\left(1+\frac2n\right),
\end{equation}
with finite size correction falling off in a power law. The above result
corresponds to a $n$-dimensional Gaussian distribution for $n$
independent components of the vector variable.
Obviously, all the values lie in the interval from
${\cal B}=0$ (Ising model, $n=1$) to ${\cal B}=2/3$ (spherical model,
$n=\infty$).

For the susceptibility, when
$tL^{1/\nu}\gg1$, one has (see Eq. (\ref{largey}))
\begin{equation}
\chi=\chi_\infty\left[1-\varepsilon\frac{n+2}{n+8}\frac{S_4}yb
L^{2-\sigma}\left(C_{4,\frac{\sigma-2}2}y^{-2}-\frac{y^{\sigma/2}}
{4S_4}\frac{\sigma+2}{\sin{\pi\sigma/2}}\right)\right].
\end{equation}
One observes that in this regime the susceptibility approaches its
bulk value not in an exponential-in-$L$, as it is usually believed
to be the case for systems with short-range critical exponents,
but in a power-in-$L$ way. The last goes beyond the standard
formulation of the finite-size scaling, but is completely
consistent with the intrinsic large-distance power-law behavior of
the correlations in systems with long-range interactions (see,
e.g. \cite{D2001} and references cited therein).

Since $\eta=O(\varepsilon^2)$ in $O(n)$ short-range models, we
were unable to verify the predicted dependence of the scaling
functions on $\eta$, which requires calculations up to second
order of $\varepsilon$, while we have retained only corrections up
to the first order in $\varepsilon$. We hope to return to this
problem in the future.

\acknowledgments
The authors thank Drs. E. Korutcheva, M. Krech, S. Romano and N. S. Tonchev
for the critical reading of the manuscript.
H. Chamati acknowledges the hospitality at the International
Centre for Theoretical Physics, Trieste.

\appendix
\section{Construction of the effective Hamiltonian}\label{effham}
Let us start with the bare Hamiltonian (\ref{model})
\begin{equation}\label{bareham}
{\cal H}\left\{\varphi\right\}=\frac12\int_V d^dx\left[
\left(\nabla\varphi\right)^2+b\left(\nabla^{\sigma/2}\varphi\right)^2
+r_0\varphi^2+\frac12u_0\varphi^4\right],
\end{equation}
where the spatial integration is over a system of linear extent
$L$ in each of its $d$ dimensions.  The partition function is
given by
\begin{equation}
{\cal Z}=\int{\cal D}\varphi\exp(-{\cal H}).
\end{equation}

Following reference \cite{rudnick85}, we spilt the field
\begin{equation}
\varphi(x)=\phi+\Sigma
\end{equation}
into a mode independent part $\phi$, which defines the
magnetization, and a part depending on the nonzero modes
$\Sigma=L^{-d} \sum_{{\bf q} \neq {\bf 0}}\varphi({\bbox
q})\exp{(i{\bbox q}\cdot {\bbox x})}$. For further calculation we
introduce the auxiliary Hamiltonian
\begin{equation}\label{haux}
{\cal H}_0\left\{\phi\right\}=L^d[\frac12r_0\phi^2+\frac14u_0\phi^4].
\end{equation}
and we treat the rest of the Hamiltonian by using perturbation
theory. Within this approximation the partition function reads
\begin{equation}\label{partfun}
{\cal Z}=\int{\cal D}\phi\exp(-H_0(\phi)-
\overset{0}{\Gamma}(\phi)),
\end{equation}
where
\begin{equation}
\overset{0}{\Gamma}=-\ln\int{\cal D}\Sigma \exp(-{\cal
H}(\phi,\Sigma)+{\cal H}_0(\phi)).
\end{equation}
Writing the difference between the bare Hamiltonian (\ref{bareham}) and the
auxiliary Hamiltonian (\ref{haux}) in the form
\begin{mathletters}
\begin{eqnarray}
{\cal H}\left\{\phi,\Sigma\right\}-{\cal H}_0\left\{\phi\right\}
&=&\frac12\int_V d^dx[(r_0+3u_0\phi^2)\Sigma^2+
\left(\nabla\Sigma\right)^2+b\left(\nabla^{\sigma/2}\Sigma\right)^2]\\
&&+\frac12 u_0\int_V d^dx[2\phi\Sigma^3+\frac12\Sigma^4].
\end{eqnarray}
\end{mathletters}
keeping in mind that the additional term involving $\int_V
d^dx\Sigma$ vanishes one gets, after some straightforward
calculations, including the evaluation of the integrals over the
field $\Sigma$,
\begin{equation}
\overset{0}{\Gamma}(\phi^2)=\frac12\sum_{{\bbox q}\neq0}\ln[r_0+{\bbox
q}^2+b{\bbox q}^\sigma]+\frac12(n+2)u_0\phi^2L^d{\cal S}_1(r_0,L)
-\frac14(n+8)u_0^2\phi^4L^d{\cal S}_2(r_0,L)+\cdots ,
\end{equation}
where
\begin{equation}\label{gamma0}
{\cal S}_m(r_0,L)=L^{-d}\sum_{{\bbox q}\neq0}\frac1{(r_0+{\bbox q}^2
+b{\bbox q}^\sigma)^m},
\end{equation}
and the dots represent terms with higher order in $\phi$.

Substituting expression (\ref{gamma0}) into that of the
the partition function (\ref{partfun}), we end up with the final
expression for the effective Hamiltonian
\begin{equation}
{\cal H}_{\rm eff.}=\frac12L^d[R\phi^2+\frac12U\phi^4],
\end{equation}
where the effective coupling constants are given by
\begin{mathletters}
\begin{eqnarray}
R&=&r_0+(n+2)u_0 L^{-d}\sum_{{\bbox q}\neq0}\frac1{r_0+{\bbox q}^2
+b|{\bbox q}|^\sigma},\\
U&=&u_0-(n+8)u_0^2L^{-d}\sum_{{\bbox q}\neq0}\frac1{\left(r_0+{\bbox q}^2
+b|{\bbox q}|^\sigma\right)^2}.
\end{eqnarray}
\end{mathletters}
These are the finite-size corrections to the bulk coupling constants
$r_0$ and $u_0$, which are necessary for the evaluation for various
thermodynamic quantities.

\section{Finite-size scaling behavior of the even moments
of the order parameter}
\label{momenta}
By definition the $2p$-th moment of the order parameter of an $O(n)$
model is given by
\begin{equation}
<M_{2p}>_n= \frac{\int_0^\infty d\Phi\Phi^{2p}
e^{-\frac{1}{2}L^d\left[ R\Phi^2+\frac{1}{2}U\Phi^4\right]}
}{\int_0^\infty d\Phi
e^{-\frac{1}{2}L^d\left[ R\Phi^2+\frac{1}{2}U\Phi^4\right]}}.
\end{equation}
Changing the variable of integration to $\varphi=(UL^d)^{1/4}\Phi$ and
by introducing the scaling variable $z=RL^{d/2}/\sqrt{U}$ the above
expression can be rewritten in the form
\begin{equation}
<M_{2p}>_n=\left(UL^d\right)^{-\frac p2}\frac{\int_0^\infty d\varphi
\varphi^{2p+n-1}e^{-\frac{1}{2} z \varphi^2-\frac{1}{4}\varphi^4}}
{\int_0^\infty d\varphi\varphi^{n-1}
e^{-\frac{1}{2} z \varphi^2-\frac{1}{4}\varphi^4}}.
\end{equation}
Using the identity \cite{RG}
\begin{equation}
\int_0^\infty x^{\nu-1}e^{-\beta x^2-\gamma
x}dx=(2\beta)^{-\nu/2}\Gamma(\nu)
\exp\left({\frac{\gamma^2}{8\beta}}\right)
D_{-\nu}\left(\frac{\gamma}{\sqrt{2\beta}}\right),
\end{equation}
where $D_{p}(z)$ are the parabolic cylinder functions,
the above expression can be rewritten in a very simple form
\begin{equation} \label{evenm}
<M_{2p}>_n=\left(UL^d/2\right)^{-\frac
p2}\frac{\Gamma[p+n/2]}{\Gamma[n/2]}
\frac{D_{-p-n/2}(z/\sqrt{2})}{D_{-n/2}(z/\sqrt{2})}.
\end{equation}
Using now the asumptotics of $D_p(z)$ \cite{RG} it is straightforward
to obtain the asumptotic behavior of the above moments for {\it i)}
$z\gg 1$  and {\it ii)} $z\ll 1$.

{\it i)} $z\gg 1$. Then one has
\begin{equation}
<M_{2p}>_n=\left(UL^d/4\right)^{-\frac
p2}\frac{\Gamma[p+n/2]}{\Gamma[n/2]}
z^{-p}\left[1-\frac{p(n+p+1)}{z^2}+O\left(\frac1{z^4}\right)\right].
\label{aszl}
\end{equation}

{\it ii)} $z\ll 1$. For this case the corresponding result is
\begin{eqnarray}
<M_{2p}>_n &=&\left(\frac{UL^d}4\right)^{-\frac p2}\frac{\Gamma[\frac
p2+\frac n4]}{\Gamma[\frac n4]}
\left[ 1+z\left(\frac{\Gamma[\frac12+\frac n4]}{\Gamma[\frac n4]}-
\frac{\Gamma[\frac p2+\frac n4+\frac12]}{\Gamma[\frac n4+\frac p2]}
\right)+ \nonumber \right. \\
& & +z^2 \left( \frac{\Gamma[\frac12+\frac n4]}{\Gamma[\frac n4]}
\left( \frac{\Gamma[\frac12+\frac n4]}{\Gamma[\frac n4]} -
\frac{\Gamma[1+\frac n4]}{2\Gamma[\frac n4+\frac12]} -
\frac{\Gamma[\frac12+\frac n4+\frac p2]}{\Gamma[\frac n4+\frac p2]} \right)
\right.\nonumber \\
& &+\left.\left.\frac{\Gamma[1+\frac n4+\frac p2]}
{2\Gamma[\frac n4+\frac p2]}\right)+ O(z^3) \right].
\end{eqnarray}
For the susceptibility ($p=1$) the above expression can be written in
the following very simple form
\begin{equation}
<M_{2}>_n  = a_n+zb_n+z^2c_n+O(z^3),
\end{equation}
where $a_n=\Gamma(\frac n4+\frac12)/\Gamma(\frac n4)$,
$b_n=a_n^2-\frac n4$, $c_n=a_n(b_n+1/4)$.

>From (\ref{aszl}) it follows that the asymptotic behavior of the
Binder cumulant is
\begin{equation}
B_n(z)=1-\frac{1}{3}\left(1+\frac{2}{n}\right)
\left[1-\frac{2}{z^2}+O\left(\frac1{z^4}\right)\right],
\end{equation}
wherefrom one has $B_n(\infty)=\frac23(1-1/n)$.


\section{Finite-size results for the physically important case: 
$\lowercase{d}+\sigma=6$}
\label{ds6}
In this Appendix we will report some results for the 
important case $d+\sigma=6$, which models the van der Waals type potential.
Note that because of the condition $d+\sigma=6$ one now  has only one 
independent variable, i.e. setting $d=4-\varepsilon$ directly leads to 
$\sigma=2+\varepsilon$. If one performs now $\varepsilon$-expansion on 
the $\sigma$-dependent terms one will in fact change the spectrum of the 
system from such one, where $q^\sigma=q^{2+\varepsilon}$ is considered as 
a perturbation to the short-range contribution (proportional to $q^2$), to 
one in which $q^\sigma$ is replaced by  $q^2+\varepsilon q^2\ln{q}$, i.e. 
where the long-range portion 
of the interaction will represent already a leading-order term. This is not 
the type of systems we are interested in. Therefore, in order to avoid this 
problem, in all the calculations below we perform $\varepsilon$-expansion 
only on the 
$d$-dependent terms and retain the full $\varepsilon$-dependence in all terms 
where it is stemming from the $\sigma$-dependence of the quantities involved. 
Following this way of acting we obtain that in the case $d+\sigma=6$ the 
expression (\ref{finalR}) for $R$ transforms into
\begin{eqnarray}\label{dsigma}
RL^2&=&y\left(1+\frac\varepsilon2\frac{n+2}{n+8}\ln y\right)-
\varepsilon\frac{n+2}{n+8}b L^{-\varepsilon} y(\ln y-2\ln L)\nonumber\\
&&+\varepsilon\frac{n+2}{n+8}S_4
\left[I^0_{\rm scaling}\left(y,4\right)-b L^{-\varepsilon}
\left(1+
y\frac\partial{\partial y}\right) I^{\frac\varepsilon2}_{\rm
scaling}\left(y,4\right)\right],
\end{eqnarray}
showing that there is an additional $\ln L$ correction to the
finite-size scaling theory. Definitely, keeping the terms
proportional to $L^{-\varepsilon}$ one goes beyond the precision
kept in the remaining part of the above equation. In accordance with the remarks made above 
note that wile one does not perform an expansion of
$L^{-\varepsilon}$ in terms of $\varepsilon$ all the terms  in
(\ref{dsigma}) proportional to $L^{-\varepsilon}$ are simply
corrections to scaling. But, once one performs that expansion,
because of the $\ln L$ proportionality, these terms produce a
leading-order contribution, which is quite unphysical. We believe
that this is an artifact of the $\varepsilon$ expansion. Such a
procedure (keeping the full $\varepsilon$-dependence in some expressions)
 has been used in reference \cite{sachdev97} in the analysis
of the scaling properties of quantum systems at low temperatures.
We hope that the above problems can be removed by performing, e.g., a field
theoretical method based on minimal renermalization at fixed space
dimensionality~\cite{chendohm2,shloms89}. This is out of the scope
of the current article.

For the coupling constant $U$, instead of (\ref{finalU}) one obtains
\begin{eqnarray}\label{udsigma}
UL^{\varepsilon}&=&\frac\varepsilon{n+8}S_4
\left[1+\frac\varepsilon2(1+\ln y)\right]-
\frac{\varepsilon^2}{n+8}bL^{-\varepsilon}S_4\left[
\ln y-2\ln L\right]\nonumber\\
&&+\frac{\varepsilon^2}{n+8}S_4^2\left[
\frac{\partial}{\partial y}I^0_{\rm
scaling}\left(y,4\right)-bL^{-\varepsilon}
\left(2\frac\partial{\partial y}+
\frac{\partial^2}{\partial y^2}\right)I^{\frac\varepsilon2}_{\rm
scaling}\left(y,4\right)\right].
\end{eqnarray}
In this limit an additional $\ln L$ correction shows up. This
expression is finite in the limit $y=0$. At the critical point it
transforms into
\begin{equation}\label{udsigma6}
UL^{\varepsilon}=\frac\varepsilon{n+8}S_4
\left[1+\frac\varepsilon2{\cal C}\right]-
\frac{\varepsilon^2}{n+8}bS_4\left[\frac12{\cal C}-2\ln L-1
\right]-\frac52\frac{\varepsilon^2}{n+8}bL^{-\varepsilon}\zeta(3).
\end{equation}
The explicit appearance of $\ln L$ will affect the result of the 
susceptibility, which will depend upon an additional $\ln L$ at the 
critical point $T=T_c$.

At the fixed point, for the `characteristic' variable $z$, we obtain
\begin{eqnarray}\label{zstar6}
z^*(y)&\equiv& \left.\frac{R L^{2}}{\sqrt{UL^\varepsilon}}
\right|_{\rm fixed point}\nonumber\\
&=&\sqrt{\frac{n+8}{\varepsilon S_4}}\left\{y-y\frac\varepsilon4\left(1-
\frac{n-4}{n+8}\ln y\right)
+\varepsilon bL^{-\varepsilon}\frac{n+2}{n+8}y+
\frac12\varepsilon bL^{-\varepsilon}\frac{4-n}{n+8}y(\ln y-2\ln L)
\right.\nonumber\\
&&+\frac{n+2}{n+8}\varepsilon S_4\left[I^0_{\rm
scaling}\left(y,4\right)
-bL^{-\varepsilon}\left(1+y\frac\partial{\partial y}\right)
I^{\frac\varepsilon2}_{\rm scaling}\left(y,4\right)\right]\nonumber\\
&&\left.-\frac12\varepsilon S_4y\left[\frac\partial{\partial
y}I^{0}_{\rm scaling}\left(y,4\right)-bL^{-\varepsilon}
\left(2\frac\partial{\partial y}+
\frac{\partial^2}{\partial y^2}\right)\right]
I^{\frac\varepsilon2}_{\rm scaling}\left(y,4\right)\right\},
\end{eqnarray}

A comparison between (\ref{zstar}) obtained for the case $d+\sigma<6$ and
(\ref{zstar6}) shows that an additional $\ln L$ appears
in the expression of the variable $z(y)$, however this does not
alter the result (\ref{zfinal}) for $z^*(0)$, i.e $z(y)$
evaluated at the critical point $T=T_c$. In this case the
term proportional to
$\ln L$ vanishes as we take the limit $y\to0$. The result 
(\ref{zstar6}) shows, in this way, that the Binder Cumulant at the critical 
point does not depend on $\ln L$ .

Far away from criticality the susceptibility (\ref{largey}) found for 
the case $d+\sigma<6$ turns into
\begin{equation}\label{largey6}
\chi=\chi_\infty\left[1+\varepsilon b\frac{n+2}{n+8}S_4^{-1}
\left(\ln\chi_\infty+5\varepsilon S_4\zeta(3) y^{-3}\right)\right].
\end{equation}
for the case $d+\sigma=6$. Remark that the susceptibility conserves the 
same features as that of the case discussed in the body paper.

\end{document}